\documentclass[12pt]{article}
\usepackage{amssymb,amsmath}
\begin{document}

\title{\textbf{Induced gravity models\\ with exact bounce solutions}}

\author{E.O.~Pozdeeva\footnote{E-mail: pozdeeva@www-hep.sinp.msu.ru} and S.Yu.~Vernov\footnote{E-mail: svernov@theory.sinp.msu.ru}\\
\small  Skobeltsyn Institute of Nuclear Physics, Lomonosov Moscow State University,\\ \small  Leninskie Gory~1, 119991, Moscow, Russia}

\date{ \ }

\maketitle

\begin{abstract}
We study dynamics of induced gravity cosmological models with the sixth degree polynomial potentials, that
have been constructed using the superpotential method.  We find conditions on the potential under which exact bounce solutions exist
and study the stability of these solutions.
\end{abstract}

\section{Introduction}

The observable evolution of the Universe~\cite{Planck2015}  can be described by the spatially flat Friedmann--Lema\^{i}tre--Robertson--Walker (FLRW) background with the interval
\begin{equation}
\label{Fried}
ds^2={}-dt^2+a^2(t)\left(dx_1^2+dx_2^2+dx_3^2\right).
\end{equation}
where $a(t)$ is the scale factor, and cosmological perturbations.

At the bounce point the period of universe contraction changes to a period of universe expansion. Thereby, a bounce point is characterized by two condition: at this point the Hubble parameter $H=\dot a/a$ is equal to zero and its cosmic time derivative $\dot H$ is positive.
 In models with standard (not phantom) scalar fields minimally coupled to gravity the Hubble parameter is monotonically decreasing function. Bounce solutions exist in models with the standard scalar field non-minimally coupled to gravity~\cite{Boisseau:2015hqa,Boisseau:2016pfh,KPTVV2015,PSTV2016}.

Models with scalar fields are very useful to describe the evolution of the FLRW metric  and play an essential role in modern cosmology. At the same time the number of integrable cosmological models is very limited~\cite{Boisseau:2015hqa,Fre,KPTVV2013}. For a generic polynomial potential cosmological models are non-integrable, moreover sometimes it is not easy to get a particular solution
in the analytic form.   Using a reconstruction procedure, one can construct such a potential of the scalar field that the resulting model with non-minimal coupling
has exact solutions with important physical properties~\cite{KTV2011,KTVV2013}. The reconstruction procedure for the models with non-minimally coupled scalar fields, proposed in~\cite{KTVV2013}, is similar to the Hamilton--Jacobi method (also known as the superpotential method or the first-order formalism) that has been applied to cosmological models with minimal coupling~\cite{SalopekBond,Chervon,AKV,Townsend,Bazeia,Andrianov:2007ua,Harko:2013gha}.

 In this paper we consider the induced gravity model with the sixth degree polynomial potential proposed in~\cite{KTVV2013}. This model has an exact solution that for some values of constants is a bounce solution~\cite{Pozdeeva2014}. We find the necessary condition of the existence and study stability of this exact  bounce solution.

\section{The superpotential method and bounce solutions}

The models with the Ricci scalar multiplied by a function of the scalar field
are described by the following action:
\begin{equation}
\label{action}
S=\int d^4 x \sqrt{-g}\left[U(\sigma)R-\frac12g^{\mu\nu}\sigma_{,\mu}\sigma_{,\nu}-V(\sigma)\right],
\end{equation}
where $U(\sigma)$ and $V(\sigma)$ are differentiable functions of the scalar field $\sigma$.

In the spatially flat FLRW universe with the interval~(\ref{Fried}),  the variation of action (\ref{action}) gives the following equations~\cite{KTVV2013,Pozdeeva2014}:
\begin{equation}
\label{Fr1}
6UH^2+6\dot U H=\frac{1}{2}\dot\sigma^2+V,
\end{equation}
\begin{equation}
\label{Fr2}
2U\left(2\dot H+3H^2\right)+4\dot U H+2\ddot U={}-\frac{1}{2}\dot\sigma^2+V,
\end{equation}
\begin{equation}
\label{Fieldequ}
\ddot \sigma+3H\dot\sigma+V^{\prime}=6\left(\dot H +2H^2\right)U^{\prime}\,,
\end{equation}
where dots mean the time derivatives and primes indicate  derivatives with respect to the scalar field
$\sigma$.
 Combining  Eqs.~(\ref{Fr1}) and (\ref{Fr2}) we obtain
\begin{equation}
\label{Fr21}
4U\dot H-2\dot U H+2\ddot U +\dot\sigma^2=0.
\end{equation}

Let $H=Y(\sigma)$ and the function ${\cal F}(\sigma)$ is defined by
\begin{equation}
\label{equsigma}
\dot \sigma={\cal F}(\sigma).
\end{equation}
Substituting $\dot\sigma$ and $\ddot \sigma={\cal F}^{\prime}{\cal F}$ into Eq.~(\ref{Fr21}), one obtains the following equation~\cite{KTVV2013}:
\begin{equation}
\label{equa}
4UY^{\prime}+2({\cal F}^{\prime}-Y)U^{\prime}+\left(2U^{\prime\prime}+1\right){\cal F}=0.
\end{equation}

The potential $V(\sigma)$ one can get from (\ref{Fr1}):
\begin{equation}
\label{potentialV}
V(\sigma)=6UY^2+6U^{\prime}{\cal F}Y-\frac{1}{2}{\cal F}^2.
\end{equation}
To find the function $\sigma(t)$ and, hence, $H(t)=Y(\sigma(t))$ we integrate Eq.~(\ref{equsigma}).

By definition a solution of Eqs.~(\ref{Fr1})--(\ref{Fieldequ}) is a bounce solution if there exists such a point~$t_b$ that
\begin{equation*}
H(t_b)=0, \qquad \dot{H}(t_b)>0.
\end{equation*}

From Eq.~(\ref{Fr1}) we get that the necessary condition for the existence of a bounce solution is $V(\sigma_b)<0$, where $\sigma_b\equiv\sigma(t_b)$. Also, from~Eq.~(\ref{Fr21}) it follows that a model with a constant positive $U$ has no bounce solutions.

If some model has been constructed by the superpotential method and we know the functions $Y(\sigma)$ and ${\cal F}(\sigma)$ explicitly, then  the search of bounce solutions is simplified, because  a value of the scalar field at a bounce point $\sigma_b$ is a solution of  the equation
$Y(\sigma)=0$.
The condition $ \dot{H}(t_b)>0$ is equivalent to $Y'(\sigma_b)F(\sigma_b)>0$.

\section{Induced gravity cosmological models with exact solutions}

In this paper, we study the induced gravity models with
$U(\sigma)= \xi\sigma^2/2$,
where $\xi$ is a positive constant. The induced gravity was first suggested by A.~Sakharov~\cite{Sakharov} and  has found many applications in cosmology~\cite{KTV2011,ind-cosm, CervantesCota,ABGV}.

Due to superpotential method, the induced gravity model with the sixth degree polynomial potential has been constructed  in~\cite{KTVV2013}. The coefficients of the potential of this model depend on three parameters. For some values of the parameters an exact bounce solution exists~\cite{Pozdeeva2014}. In this paper we continue the consideration of this  model and study   conditions for existence of bounce  solutions and their  behavior.

Let $Y(\sigma)$ is a generic quadratic polynomial
$Y(\sigma)=C_0+C_1\sigma+ C_2\sigma^2$,
where $C_0$, $C_1$, and $C_2$ are arbitrary constants, but $C_0\neq 0$ and $C_2\neq 0$. From  Eq.~(\ref{equa}) we obtain
\begin{equation}
{\cal F}(\sigma)=\frac{2\left[(8\xi+1)C_0-(4\xi+1)C_2\sigma^2\right]\xi\sigma}{(4\xi+1)(8\xi+1)}+ B\sigma^{-(1+2\xi)/(2\xi)},
\end{equation}
where $B$ is an arbitrary constant.
 When $B=0$, the function ${\cal F}(\sigma)$ is a cubic polynomial and  the general solution for Eq.~(\ref{equsigma}) can be written in terms of elementary functions~\cite{KTVV2013}:
\begin{equation}
\label{stc}
\sigma_\pm(t)=\pm\frac{\sqrt{(8\xi+1)C_0}}{\sqrt{(8\xi+1)C_0ce^{-\omega t}+(4\xi+1)C_2}}\,,
\end{equation}
where $\omega=4\xi C_0/(4\xi+1)$, $c$ is an integration constant.
The function  $\sigma_\pm$ should be real at any moment of time, therefore, considering limits at $t\rightarrow\pm\infty$ we get two possibilities:  $C_0>0$ and $C_2>0$,  or $C_0<0$ and $C_2<0$. In both cases $c>0$.
If $C_0<0$ then the Hubble parameter tends to $C_0$  at  $t\rightarrow+\infty$, that contradicts to the observable expansion of the Universe at late times. By this reason we restrict ourself to the case $C_0>0$.

The potential of the model considered is the sixth degree polynomial~\cite{KTVV2013,Pozdeeva2014}:
\begin{equation}
\label{V6}
\begin{split}
V(\sigma)&=\frac{(16\xi+3)(6\xi+1)\xi}{(8\xi+1)^2}C_2^2\sigma^6+\frac{6(6\xi+1)\xi}{8\xi+1}C_1C_2\sigma^5+\frac{6(6\xi+1)\xi}{4\xi+1}C_0C_1\sigma^3
+{}\\
&
{}+\left[3\xi C_1^2+\frac{2(6\xi+1)(20\xi+3)\xi}{(8\xi+1)(4\xi+1)}C_0C_2\right]\sigma^4+\frac{(16\xi+3)(6\xi+1)\xi}{(4\xi+1)^2}C_0^2\sigma^2\,.
\end{split}
\end{equation}
The change of sings both $C_1$ and $\sigma$ does not change the value of the potential and the Hubble parameter. So, we can consider  the solutions $\sigma_+$ only. Note that for all $t$ $\sigma_+(t)>0$.
All obtained results will be correct for $\sigma_-$ and the potential with $-C_1$ as well.

\section{Existence and stability of exact bounce solutions}

Let us find conditions that are necessary for the existence of a bounce solution.
The first restriction on parameters $C_i$ we get from the equation $Y=0$ that has the following solutions:
\begin{equation}\label{sigmab}
    \sigma_{b\pm}=\frac{-C_1\pm\sqrt{C_1^2-4C_0C_2}}{2C_2}.
\end{equation}
These solutions are real solution only if
$C_1^2\geqslant{}4C_0C_2$.
We consider the  $C_0>0$, $C_2>0$, and $\sigma_+>0$, so, the model get the bounce only at $\sigma=\sigma_{b+}>0$ and only under condition $C_1<{}- 2\sqrt{C_0C_2}$.
Note that the value of the potential at the bounce point  does not depend on $C_1$, because
$V(\sigma_b)=-{\cal F}^2(\sigma_b)/2$.

All  considering exact solutions tend to de Sitter ones. In our paper~\cite{Pozdeeva2014} it has been found that de Sitters solutions
are unstable at
\begin{equation}
\label{condnonmon}
 C_1 <{}-\frac{2(16\xi+3)}{3\sqrt{(8\xi+1)(4\xi+1)}}\sqrt{C_0C_2}.
\end{equation}
The function
$\frac{2(16\xi+3)}{3\sqrt{(8\xi+1)(4\xi+1)}}$
is a monotonically decreasing function that is equal to 2 at $\xi=0$. So, we come to conclusion that at any $\xi>0$ all bounce exact solutions $\sigma_+$ tend to unstable de Sitter solutions.   The corresponding Hubble parameter is a monotonically increasing function after a bounce, because $\dot H(\sigma)>\sqrt{C^2_2-4C_0C_2}\dot\sigma>0$ for all $\sigma>\sigma_{b+}$. Surely, such solutions can not describe the evolution of the observable Universe, because inflation corresponds to a decreasing Hubble parameter, but these solutions are not unique bounce solutions in the model considered. Using the symmetry of the potential with respect to the change $\sigma_+$ on $\sigma_-$ and $C_1$ on $-C_1$, we come to conclusion that any exact bounce solution tends to unstable de Sitter ones.

\section{Conclusion}

We have found necessary condition of existence of the exact bounce solutions that has been constructed using superpotential method. All bounce solutions that can be presented in the analytic form (\ref{stc}) tends to unstable de Sitter solution.

The exact bounce solutions of the considering model are not able to describe the evolution of observable Universe. It does not mean that  they are useless, because it can be possible to slightly  modified these solutions to do not loss the bounce points but get more suitable behaviour after this point. It demand numerical calculations and maybe some modification of the potential.  Note that exact bounce solutions obtained in the integrable cosmological model~\cite{Boisseau:2015hqa} corresponds to monotonically increasing Hubble parameter, whereas as  slightly modified models that are not integrable~\cite{Boisseau:2016pfh,PSTV2016} allow to get bounce solutions with non-monotonic Hubble parameter. We hope that further disquisition with numeric calculations gives bounce solutions with an interesting non-monotonic behaviour of the Hubble parameter in  this model or in a slightly modified model.

This work was partially supported by grant NSh-7989.2016.2 of the President of Russian Federation.
Research of E.P. is supported in part by grant MK-7835.2016.2  of the President of Russian Federation.

\end{document}